\documentclass[%
 aip,
 amsmath,amssymb,
 reprint,
]{revtex4-1}
\usepackage[usenames,dvipsnames]{color}
\usepackage{soul}
\usepackage{xcolor}
\usepackage{color,soul}
\usepackage{graphicx}
\usepackage{dcolumn}
\usepackage{bm}
\usepackage[colorlinks = true,
            linkcolor = blue,
            urlcolor  = blue,
            citecolor = blue,
            anchorcolor = blue]{hyperref}
\hypersetup{urlcolor=blue, colorlinks=true}  
\usepackage{etoolbox}
\usepackage{filecontents}
\usepackage[titletoc,title]{appendix}

\newcommand{\ga}{\gamma}
\newcommand{\sig}{\sigma}
\newcommand{\al}{\alpha}

\begin{document}
\title{Effect of pseudospin polarization on wave packet dynamics in  graphene antidot lattices (GALs) in  the presence of a normal magnetic field }
\author{R. A. W. Ayyubi}
\affiliation{Department of Physics, Quaid-I-Azam University, Islamabad 45320, Pakistan}
\author{N. J. M. Horing}
\affiliation{Department of Physics, Stevens Institute of Technology, Hoboken, New Jersey 07030, USA}
\author{K. Sabeeh}
\affiliation{Department of Physics, Quaid-I-Azam University, Islamabad 45320, Pakistan}
\begin{abstract}
We have investigated the role of pseudospin polarization in electron wave packet dynamics in pristine graphene and in a graphene antidot lattice subject to an external magnetic field. Employing a Green's function formalism, we show that the electron dynamics can be controlled by tuning pseudospin polarization.
We find that in Landau quantized pristine graphene both the propagation of an electron wave packet and Zitterbewegung oscillations strongly depend on pseudospin polarization. The electron wave packet is found to propagate in the direction of initial pseudospin polarization. We also show that, in this system, the propagation of an electron can be enhanced in any desired direction by carving a one dimensional antidot lattice in that direction. The study suggests that a graphene antidot lattice can serve as a channel for electron transport with the possibility of tunability by means of  pseudospin polarization, antidot potential and applied normal magnetic field strength. 

\end{abstract}
\maketitle
\section{\label{sec:level1}INTRODUCTION}
It has been discovered that the charge carriers in two-dimensional materials such as graphene have, in addition to spin, other binary degrees of freedom. These are sublattice pseudospin and valley pseudospin.\cite{articlepss, PhysRevLett.106.116803} The graphene honeycomb lattice is composed of two triangular sublattices; this allows representing the  wave function of electrons as a two-component spinor, like a spin-$1/2$ particle.\cite{PhysRevLett.106.116803} Similarly extrema in the band structure at $K$ and $K^\prime$ points also define a binary degree of freedom which is known as the valley pseudospin. The established field of spintronics exploits spin for electronic applications.\cite{naturearticle, PhysRevLett.102.247204, PS2, Kamalakar2015LongDS, PhysRevA.57.120, BSA} Efforts to understand and control these additional degrees of freedom with the aim of utilizing them in technological applications has led to the emerging fields of  pseudospintronics and valleytronics.\cite{10.1021/nn300997f, PhysRevLett.116.106601} Towards this aim, in this work we focus on the sublattice pseudospin degree of freedom of the prototypical 2D material, which is graphene. Specifically, we address the role of pseudospin polarization in the propagation of electrons in graphene in the presence of both an applied magnetic field and an antidot lattice.
\\
We also investigate the role of pseudospin polarization on Zitterbewegung (ZB), which is the oscillatory motion of the wave packet, another important phenomenon observed in Dirac materials.\cite{RevModPhys.81.109} These oscillations have features which differ in the presence and in the absence of a magnetic field.  In monolayer graphene, in the absence of a magnetic field, ZB oscillations  are transient  and disappear after some time, while  for finite B, ZB oscillations are  permanent and do not die with time. \cite{PhysRevB.78.235321, rusinZBinB2008} A recent study suggests that ZB oscillations in monolayer graphene in the absence of magnetic field strongly depends on initial pseudospin polarization, which is named as pseudospinorial Zitterbewegung (PZB). \cite{PZB} These studies invite us to explore the effect of initial pseudospin polarization on ZB  in graphene in the presence of a magnetic field, which we have investigated in this study.
\\
Our main finding is that the  interplay of  pseudospin polarization, applied magnetic field and antidot lattice leads to enhanced propagation in the direction of the antidot lattice; this raises the possibility of creating one-dimensional channels for electrons in Landau quantized graphene. The control features for propagation in  these channels are the applied magnetic field, strength of antidot lattice and pseudospin polarization. Further, we also show that there are persistent Zitterbewegung (ZB) Oscillations in the presence of an applied magnetic field in pristine graphene, which are sensitive to pseudospin polarization.
\section{\label{sec:level1}Graphene in A Perpendicular Magnetic Field: GREEN'S FUNCTION}
Graphene is a single layer of carbon atoms packed in a honeycomb lattice. The effective Hamiltonian of  graphene has the form \cite{wallace1947}
\begin{equation} \label{eq:p1}
H=\ga \hspace{0.6mm}\boldsymbol{\sig}_{\nu} \hspace{0.6mm}  . \hspace{0.6mm}\bf{ p}
\end{equation}
where $\sig_{\nu}=[\sig\textsubscript x,1_\nu  \hspace{1mm}\sig\textsubscript y]$ and  $\sig\textsubscript x$, $\sig\textsubscript y$ are Pauli's spin matrices, which act on the sublattice/pseudospin space and represent the sublattice degree of freedom of graphene's honeycomb lattice structure, 
also $1_\nu=1$ or -1 for $K$ or $K^\prime$ valleys,   and $\ga=\frac{3}{2} \al_h \textit{d}\approx 10^{6}ms^{-1}$ plays the role of a density-independent Fermi velocity ($\al_h$ is the  hopping amplitude originating from the tight binding approximation in which the lattice spacing is \textit{d}).
\\
We consider a graphene sheet placed on the $xy$-plane in a perpendicular and uniform magnetic field $\textbf B$= $B\hat z$ with vector potential $\textbf A = \frac{1}{2}  (\textbf B \times \textbf r)$. The magnetic field is introduced by minimal substitution $\textbf p\rightarrow \textbf p - e \textbf A$ in Eq. (\ref{eq:p1}).
 The requirement of gauge invariance leads to 
\begin{equation} \label{eq:p18} 
\mathcal G(\textbf r, \textbf r^\prime;t,  t^\prime)= C(\textbf r,\textbf r^\prime) \hspace{1mm} \mathcal G^\prime(\textbf r - \textbf r^\prime;t - t^\prime),
\end{equation}
 where the Green's function  $\mathcal G^\prime(\textbf r - \textbf r^\prime;t - t^\prime)$ is spatially translationally invariant and gauge invariant and it is related to usual Green's function $\mathcal G(\textbf r, \textbf r^\prime;t,  t^\prime)$ with the help of  Peierls phase factor $C(\textbf r,\textbf r^\prime)$, which carries all aspects of the lack of translational invariance in a magnetic field and all gauge dependence as
\begin{equation} \label{eq:p7}
C(\textbf r,\textbf r^\prime)=exp\left( \frac{\textit i e}{2 \hbar }\textbf r \hspace{0.5mm} \cdot \hspace{0.5mm} \textbf B \times \textbf r^\prime -\phi(\textbf r)+\phi(\textbf r^\prime)\right)
\end{equation}
($\phi(\textbf r)$ is an arbitrary gauge function). The translationally invariant Green's function is given by the equation of motion
\begin{multline} \label{eq:p50}
\left[\textit{i} \frac{\partial}{\partial T} - \ga \sig_{\nu} \cdot \left(\frac{1}{\textit{i}} \frac{\partial}{\partial \textbf R} - \frac{e}{2} \textbf B \times \textbf R\right)\right] \mathcal G^\prime (\textbf R, T) =
\\
  I\textsubscript 2 \delta\textsuperscript 2 (\textbf R) \delta (T)
\end{multline}
where  $\textbf R =\textbf r - \textbf r^\prime$ such that $X=x-x^\prime, Y=y-y^\prime$ and $T=t-t^\prime$.
\\
The diagonal elements  of the Green's function matrix ($\omega$ representation) for this system of  monolayer graphene in a uniform, constant, perpendicular magnetic field are   ($\mathcal G^\prime_{11\atop 22}$ collectively represents $\mathcal G^\prime_{11}$, $\mathcal G^\prime_{ 22}$) 
\begin{multline}\label{eq:p2}
\mathcal G^\prime_{11\atop 22}(\textbf R;\omega)= \frac{1}{4\pi\hbar\ga^2}\hspace{0.8mm} \omega \hspace{0.8mm} e^{-\frac{\omega^2_g}{8\ga^2}\left(X^2+Y^2\right)}  
\\
\times \sum_{n=0}^\infty
 \frac{ L_n\left[\frac{\omega^2_g}{4\ga^2}\left(X^2+Y^2\right)\right]}{\frac{\omega^2}{\omega^2_g}-\left(n+\frac{1 - 1_\nu}{2}\right)} .
\end{multline} 
We introduced the notation
\begin{equation} \label{eq:p3}
\omega_g=\ga\sqrt{\frac{2eB}{\hbar}},
\end{equation}
which is the cyclotron frequency for  Dirac fermions. The poles of $\mathcal G^\prime_{11\atop 22}(\textbf R;\omega)$  show that the energy spectrum is Landau quantized, with  Landau level index  given by $n$.
\\
The off-diagonal matrix elements can  be calculated using the relation  ($\ga_\nu\equiv1_\nu\ga$) \cite{horing2009}

\begin{equation} \label{eq:p4}
\omega \mathcal G^\prime_{21\atop12}= [\ga \Pi\textsubscript X \textsubscript Y \pm \textit{i}\ga_{\nu}  \Pi\textsubscript Y \textsubscript X] \mathcal G^\prime_{11\atop22}.
\end{equation}
In the above equation, $ \Pi\textsubscript X \textsubscript Y  \equiv \frac{1}{\textit{i}} \frac{\partial}{\partial X} + \frac{eB}{2} Y $  and $\Pi\textsubscript Y \textsubscript X \equiv \frac{1}{\textit{i}} \frac{\partial}{\partial Y} - \frac{eB}{2} X $ are momentum operators and $\mathcal G^\prime_{21 \atop 12}$= $\mathcal G^\prime_{21}$ or $\mathcal G^\prime_{12}$ corresponds to the upper or lower $\pm$, $\mp$ signs elsewhere in the equations. 
Making use of Eq. (\ref{eq:p4}), off-diagonal elements of Green's function matrix  for $K$ and $K^\prime$ can be separately expressed as
\begin{multline}\label{eq:p5}
K(1_\nu=+1) :  \hspace{1.2mm}   \mathcal G^\prime_{21\atop 12}(\textbf R;\omega)= \frac{\omega_g^2}{8\pi\hbar\ga^3} \hspace{0.8mm} e^{-\frac{\omega^2_g}{8\ga^2}\left(X^2+Y^2\right)}  
\\
\times \left(\textit i X \mp Y\right) \sum_{n=1}^\infty
 \frac{ L_{n-1}^1 \left[\frac{\omega^2_g}{4\ga^2}\left(X^2+Y^2\right)\right]}{\frac{\omega^2}{\omega^2_g}-n}
\end{multline}
\begin{multline}\label{eq:p6}
K^\prime (1_\nu=-1) :  \hspace{1.2mm}   \mathcal G^\prime_{21\atop 12}(\textbf R;\omega)= \frac{\omega_g^2}{8\pi\hbar\ga^3}  \hspace{0.8mm} e^{-\frac{\omega^2_g}{8\ga^2}\left(X^2+Y^2\right)}  
\\
 \times \left(\textit i X \pm Y\right) \sum_{n=1}^\infty
 \frac{ L_{n-1}^1 \left[\frac{\omega^2_g}{4\ga^2}\left(X^2+Y^2\right)\right]}{\frac{\omega^2}{\omega^2_g}-\left(n+1\right)}.
\end{multline}
 Clearly, the off-diagonal elements interchange with the interchange of Dirac points ($K$ and $K^\prime$). \cite{horinglandau2017, addhoring2018}  
\\
The Green's functions matrix  in time  representation can be obtained by Fourier  transform of  Eqs.  (\ref{eq:p2}), (\ref{eq:p5}), (\ref{eq:p6}) as
\begin{equation}  \label{eq:p8}
\mathcal G^\prime_{\mu\nu}(\textbf R; t) = \int_{-\infty}^\infty \frac{ d\omega}{2\pi} \hspace{0.8mm} e^{-\textit i \omega t} \hspace{0.8mm} \mathcal G^\prime_{\mu\nu}(\textbf R;  \omega) 
\end{equation}
where $\mu , \nu = 1 , 2$ denote matrix indices.
\\
Noting  that $\mathcal G^\prime_{\mu\nu}(\textbf R; \omega)$ has real energy poles at $\epsilon_K=\pm \omega_g \hspace{1mm}\sqrt{n}$ ; $\epsilon_{K^\prime}=\pm \omega_g \hspace{1mm}\sqrt{n+1}$  we employ contour integration with $\omega\rightarrow\omega+\textit i\hspace{0.6mm}0^+$ for the retarded Green's function with  the contour closed in the lower half plane  running clockwise from $-\infty$ to $+\infty$. For the $K$-point, we have (for the $K^\prime$-point $\sqrt n\rightarrow \sqrt{n+1}$)
\begin{multline}  \label{eq:p9}
\mathcal G^\prime_{11 \atop 22}\left( x,x^\prime;y,y^\prime;t\right)=-\textit{i} \hspace{0.7mm} \eta_+(t) \hspace{0.7mm} \frac{  \omega_g^2}{4\pi \hspace{0.5mm}\gamma^2} \hspace{1.2 mm}  e^{-\frac{1}{2}\hspace{0.7 mm}\zeta} \hspace{1.2 mm} \sum_{n=0}^\infty
 L_n\left[ \zeta \right]  \hspace{1mm}
\\
\times \cos \left(\frac{\omega_g \hspace{0.8mm} l}{\ga} \hspace{0.5mm} \frac{t}{\tau_o} \hspace{0.4mm} \sqrt n \hspace{0.8mm}\right) 
\end{multline}

\begin{multline}  \label{eq:p10}
\mathcal G^\prime_{21 \atop 12}\left( x,x^\prime;y,y^\prime;t\right)=- \eta_+(t) \hspace{0.7mm} \frac{ \omega_g^3 \hspace{0.8mm} l}{8\pi \hspace{0.5mm}\gamma^3} \hspace{1.2 mm}  e^{-\frac{1}{2}\hspace{0.7 mm}\zeta} \hspace{1.2 mm}
\\
\times \left[\textit i \left(\frac{x-x^\prime}{\textit l}\right)\mp \left(\frac{y-y^\prime}{\textit l}\right)  \right]
 \sum_{n=1}^\infty \frac{ L_{n-1}^1\left[ \zeta \right] }{\sqrt n}
\ \hspace{1mm}
\\
\times \sin \left(\frac{\omega_g \hspace{0.8mm} l}{\ga} \hspace{0.5mm} \frac{t}{\tau_o} \hspace{0.4mm} \sqrt n \hspace{0.8mm}\right) :
\end{multline}
Here, we have introduced an arbitrary constant length, $l$, chosen for convenience to be the width of an impressed wave packet, and
 $\eta_+(t)=0,1$ for $t\leq0,  t\textgreater 1$, respectively, is the Heaviside unit step function; also, we have defined
\begin{equation*} \label{eq:p11}
\zeta =  \frac{1}{4}\left(\frac{\omega_g \hspace{0.8mm} \textit l}{\gamma}\right)^2 \left[\left(\frac{x-x^\prime}{\textit l}\right)^2+\left(\frac{y-y^\prime}{\textit l}\right)^2 \right] 
\end{equation*}
 and $\tau_o = \textit l/\ga$.
 \\
To study the wave packet dynamics, we take the initial wave function to be a Gaussian wave packet having nonvanishing average momentum $p_{0x} = \hbar k_{0x}$ and width $l$,
\begin{equation} \label{eq:p13}
\psi(\textbf r,0) =  \frac{f(\textbf r) }{\sqrt{\mid c_1\mid ^2+\mid  c_2 \mid ^2}} 
\begin{pmatrix} 
c_1  \\
c_2  
\end{pmatrix}
\end{equation}
\begin{equation*} \label{eq:p14}
f(\textbf r) = \frac{1}{l  \hspace{0.4mm} \sqrt\pi }  \hspace{0.5 mm}  exp\left(-\frac{x^2+y^2}{2\hspace{0.4mm} l^2}+\textit i k_{0x} x\right) ,
\end{equation*}
where  $c_1$ and $c_2$  are the coefficients which set the initial pseudospin polarization. Also $\psi(\textbf r,0)$ can be taken as a smooth enveloping function considering  that lattice period is much smaller than the   width $l$ of  the initial wave packet. The
 Green's function matrix elements $\mathcal G_{\mu\nu}(\textbf r, \textbf r^\prime,t)$   determine the time evolution of an arbitrary initial state $\psi(\textbf r,0)$;  in Schr$\ddot{o}$dinger representation it  is given by
\\ 
\begin{equation} \label{eq:p15}
\psi_\mu(\textbf r,t) = \int  d\textbf r^\prime  \hspace{0.8 mm} \mathcal G_{\mu\nu}(\textbf r, \textbf r^\prime,t) \hspace{0.8 mm} \psi_\nu (\textbf r^\prime,0)  \hspace{1.2 mm},
\end{equation}
where $\mu, \nu = 1,2$ denote the matrix indices, which corresponds to upper component $\psi_1(\textbf r,t)$ and lower component $\psi_2(\textbf r,t)$  of state $\psi_\mu(\textbf r,t)$.
 The probability density will be
\begin{equation} \label{eq:p16}
\rho(\textbf r,t) =  \mid\psi(\textbf r,t)\mid^2 =  \mid\psi_1(\textbf r,t)\mid^2 +\mid\psi_2(\textbf r,t)\mid^2 .
\end{equation}
And to study ZB, the average value of coordinates  can be represented as
\begin{equation}  \label{eq:p17}
\bar x_j = \int \psi_1^\ast(\textbf r,t)  \hspace{0.4mm}  x_j \hspace{0.4mm} \psi_1(\textbf r,t) \hspace{0.4mm}  d\textbf r +  \int \psi_2^\ast(\textbf r,t) \hspace{0.4mm}  x_j \hspace{0.4mm} \psi_2(\textbf r,t) \hspace{0.4mm} d\textbf r
\end{equation}
where $j=1,2$  with $x_1$=$x$ and $x_2$=$y$.
The Peierls phase factor defined in Eq. (\ref{eq:p7})  for the choice $\phi(\textbf r)=\phi(\textbf r^\prime)\equiv0$  has the form
\begin{equation*} \label{eq:p20}
C(\textbf r,\textbf r^\prime)=exp\left[ \frac{\textit i }{4 } \left(\frac{\omega_g l}{\ga}\right)^2 \left(\frac{y x^\prime}{l^2}-\frac{x y^\prime}{l^2} \right) \right] .
\end{equation*}
\section{\label{sec:level1}Dynamics of A Gaussian wave packet with different pseudospin polarizations in Landau quantized graphene: Zitterbewegung}
\begin{figure}[t!]
\centering
\includegraphics[scale=0.68]{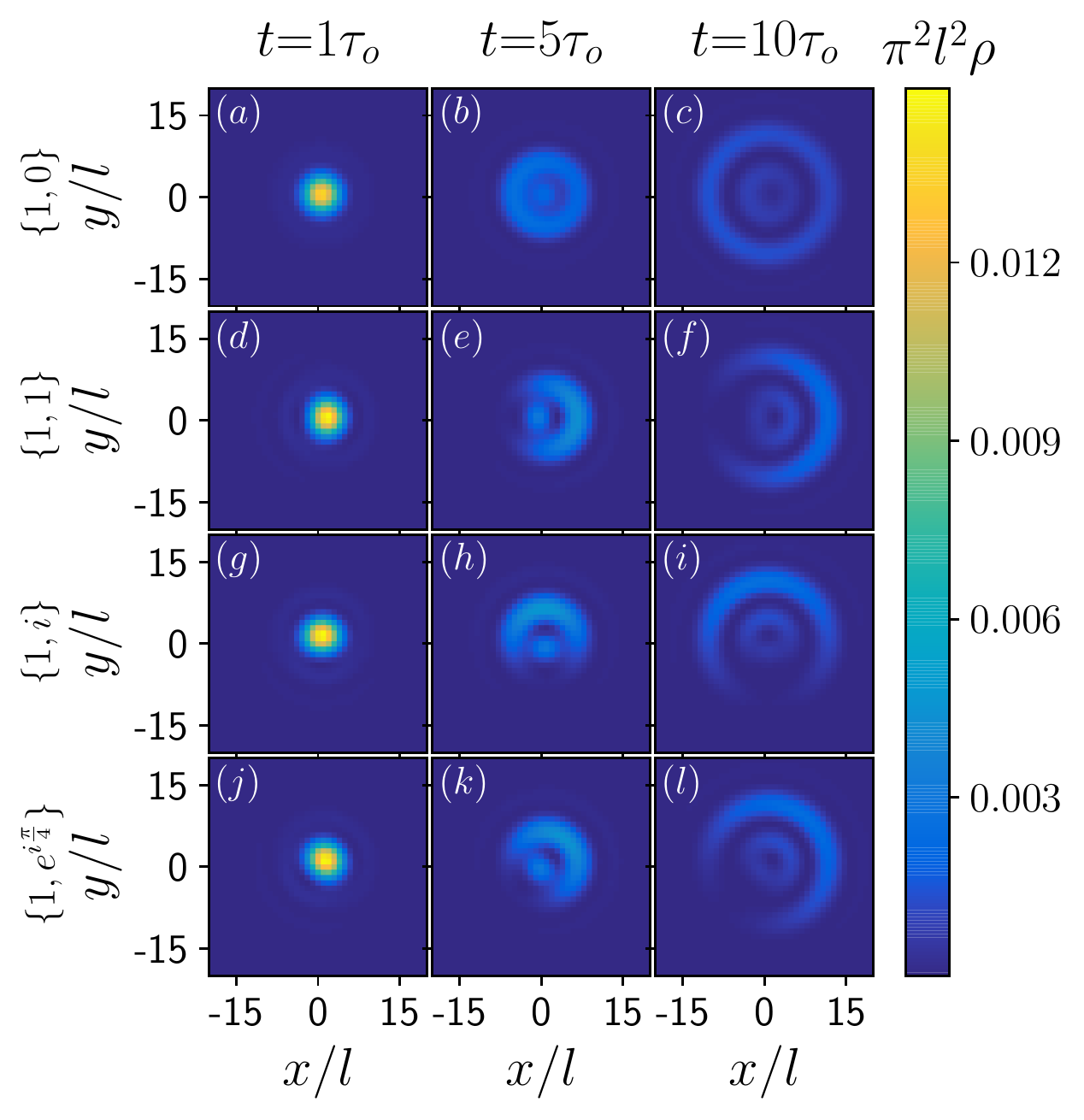}  
\caption{ (Color online) The electron probability density $\rho(x,y,t)$ for initial Gaussian wave packet with $B=3.3 \hspace{1mm} Tesla$, $a_0$=$k_{0x}\hspace{0.8mm}l$=$1.2$ at times $1\tau_o$, $5 \hspace{0.5mm}\tau_o$ and $10\hspace{0.5mm}\tau_o$. Time increases as we move from left to right. Color bar given at extreme right side shows the strength of probability density from its minimum to  its maximum. Up to down: Four rows with initial pseudospin polarizations 
$\{1,0\}$,$\{1,1\}$,$\{1,i\}$ and $\{1,e^{i\frac{\pi}{4}}\}$ with pseudospin and propagation directed in the radial, $x$, $y$ and $(x=1,y=1)$  directions respectively.}
\label{Fig-1} 
\end{figure}

To obtain  results for the temporal evolution of the initial Gaussian wave packet, ZB oscillations and the effect of  initial-pseudospin polarization, we have performed  numerical calculations. To facilitate it, we have introduced following dimensionless variables: \cite{PhysRevB.78.235321, wpdskii2008}
\begin{itemize}
\item{A dimensionless parameter which is suitable to replace wave vector $k_{0x}$ is  $a_0=k_{0x} \hspace{0.3 mm} l $. }
\item{Distance for propagation of the wave packet can be measured in  units of initial width  $ l $ of wave packet.}
\item{ Time can be measured in $\tau_o$=$ l/\ga $  ($\ga$ is Fermi velocity, $10^6$ m/s) units.}
\item{Some other variables can be combined to produce  dimensionless variables, e.g
$b= \frac{\omega_g \hspace{0.6mm}l}{\ga}$.}
\end{itemize}
The Landau level summation is performed up to the $10th$ Landau level in all calculations since the results become  convergent in this limit.
\\
We  consider four cases with different initial pseudospin polarizations $\{c_1,c_2\}$ for the Gaussian wave packet given in Eq. (\ref{eq:p13}).
\\
Case-1: $\{c_1,c_2\}$=$\{1,0\}$ which corresponds to initial electron probability of  one at the sites of sublattice A. 
\\
Case-2:$\{c_1,c_2\}$=$\{1,1\}$  corresponds to the situation where the pseudospin is directed along the $x$-axis.
\\
Case-3: $\{c_1,c_2\}$=$\{1,i\}$ \hspace{0.5mm} this corresponds to  the pseudospin  directed along $y$-axis.
\\
Case-4: $\{c_1,c_2\}$=$\{1,e^{\textit i\pi/4}\}$ \hspace{0.5mm} this implies that at  time   $t$=$0$, the pseudospin  lies in $x$-$y$ plane making an angle of $45^o$ with $x$-axis.
\\
The numerical results obtained from  Eqs. (\ref{eq:p15}), (\ref{eq:p16})  are plotted in Fig.\ref{Fig-1}. \\
In Fig.\ref{Fig-1}, electronic probability density is plotted for parameters $t$=$1, 5, 10  \hspace{0.8mm} \tau_o$, with momentum  $k_{0x} = 0.6 \hspace{0.8mm} nm^{-1}$,  width $l$=$2\hspace{0.8mm} nm$  so that $a_0$=$k_{0x}\hspace{0.8mm}l$=$1.2$ and $\tau_o$=$2\hspace{0.8mm} femtosecond$ and $B$=$3.3 \hspace{0.8mm}$ in units of $  Tesla$ i.e. $\omega_g \simeq 1\times10^{14} \hspace{0.8mm} Hz$.  $B$=$3.3 \hspace{0.8mm}T$ is chosen to facilitate comparison with work in the literature and also because it facilitate numerical computation as $\omega_g$ is a round figure at this value. \cite{PhysRevB.78.235321} 
Left panel shows the initial wave packet at a very small time $t$=$1  \hspace{0.8mm} \tau_o$, and as we move from left to right the time evolution of initial Gaussian packet can be seen for different initial pseudospin polarizations. The strength of the electron probability density $\rho(x,y,t)$  is given by the color bar on the right side. 
\\
As one can see in Fig.\ref{Fig-1} $(a),(b),(c)$, the wave packet spreads and propagates in the plane of graphene sheet in the form of rings for the pseudospin $c_1$=$1$ and $c_2$=$0$. Initially in Fig.\ref{Fig-1} $(a)$, maximum probability of the electron is located at the origin, but as  time increases, the wave packet propagates and the electron density can be found at a radius of $r\simeq24 \hspace{0.8mm} nm$ in $20 \hspace{0.8mm} fs$.
 Similarly for other three cases, the wave packet propagates with its maximum probability density in the direction of initial  pseudospin polarizations, but in the shape of incomplete rings, this is because the probability density gradually decreases in the directions away from the direction of pseudospin polarization. It propagates in the $x$ direction when initial  pseudospin polarization is along $x$ axis (see Figs.\ref{Fig-1} $(d),(e),(f)$), in the $y$ direction when initial  pseudospin polarization is along $y$ axis (see Figs.\ref{Fig-1} $(g),(h),(i)$)  and it propagates in  $(x,y)$=$(1,1)$ direction when initial pseudospin is polarized  along  $(x,y)$=$(1,1)$ direction (see Figs.\ref{Fig-1} $(j),(k),(l)$). This is because of the conservation of chirality, in which momentum gets aligned with pseudospin and  $\vec \sig_{\nu} \hspace{0.6mm}  . \hspace{0.6mm}\vec p$ remains conserved. Also the distance covered by the wave packet in these three cases is $r\simeq24 \hspace{0.8mm} nm$ in $20 \hspace{0.8mm} fs$.  Hence the direction of propagation of a wave packet in Landau quantized graphene can be controlled using pseudospin polarization. Also, the wave packet propagates without any splitting; splitting was observed  in the case of  monolayer graphene in the absence of magnetic field.\cite{PhysRevB.78.235321} 
We propose that, experimentally, this type of  controlled propagation of a wave packet in any direction can easily be obtained using photonic graphene as test  beds.\cite{PS2}
\\
 Also  note that the electronic probability densities plotted in Fig.\ref{Fig-1}   are not symmetric  with respect to both  $x$ and $y$ axes:  $\rho(x,y,t)_{\{c_1,c_2\}}\neq \rho(-x,y,t)_{\{c_1,c_2\}}$ and $\rho(x,y,t)_{\{c_1,c_2\}}\neq \rho(x,-y,t)_{\{c_1,c_2\}}$ for any pseudospin polarization (subscript $\{c_1,c_2\}$ defines the corresponding pseudospin polarization): This means that center of the  wave packet is oscillating along both the $x$ as well as the $y$ directions; these oscillations can be readily recognized as  Zitterbewegung oscillations. \\
\begin{figure}[t]
\centering
\includegraphics[scale=0.53]{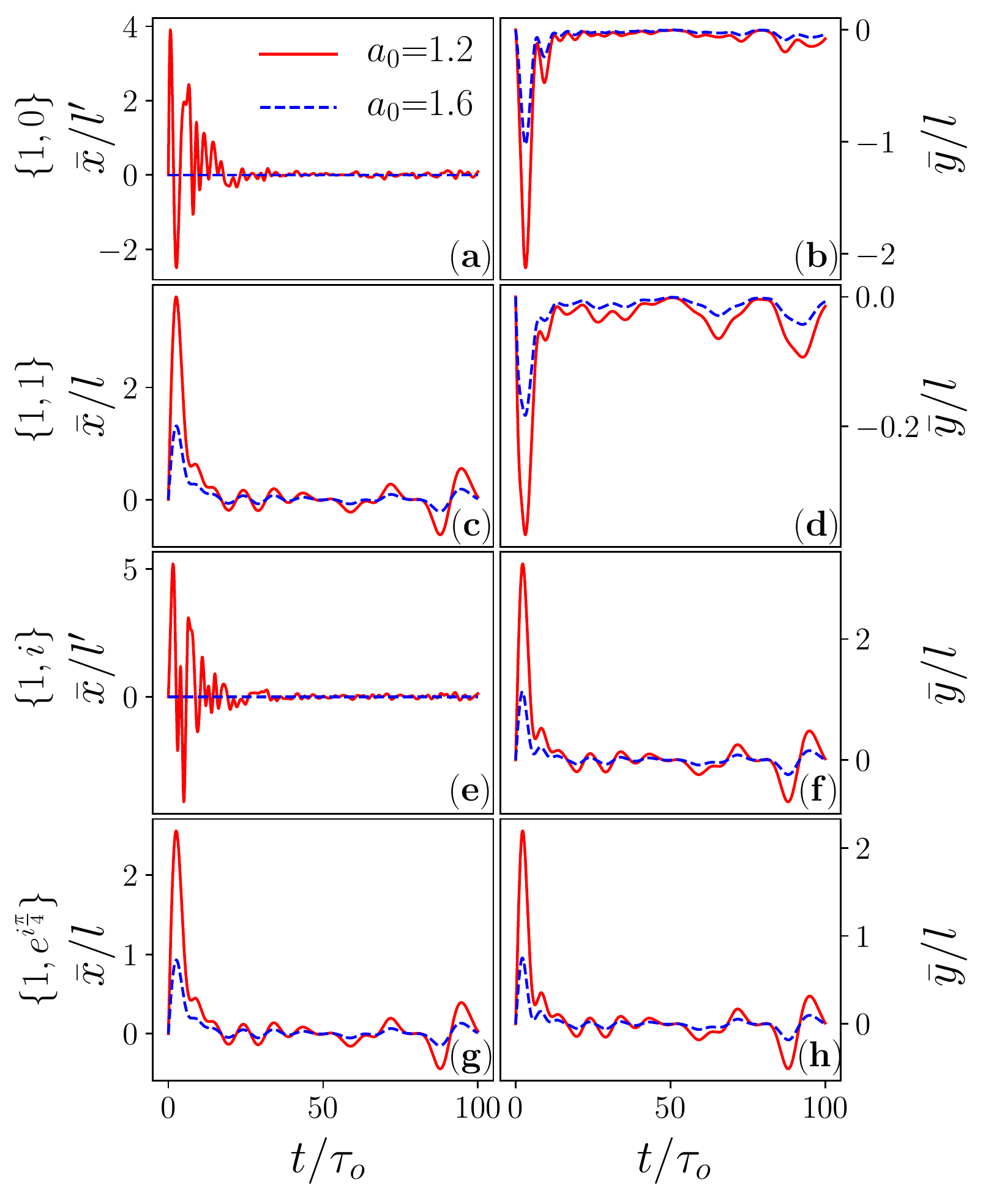}
\caption{Average coordinates $\bar x(t)$ (left column) and $\bar y(t)$ (right column) versus time $(\tau_o=l/\gamma)$ corresponding to four different values of pseudospin polarization (changing from up to down), at two values of momentum $a_0$ with $B=3.3 \hspace{1mm} Tesla$. (a),(c),(e),(g): $\bar x(t)$ versus time corresponding to pseudospin polarizations $\{1,0\}$,$\{1,1\}$, $\{1,i\}$ and $\{ 1 , e^{i\frac{\pi}{4}}\} $ respectively. (b),(d),(f),(h): $\bar y(t)$ versus time corresponding to pseudospin polarizations $\{1,0\}$,$\{1,1\}$, $\{1,i\}$ and $\{ 1 , e^{i\frac{\pi}{4}}\} $ respectively. Here $l^\prime$=$l\times10^{-5}$. }
\label{Fig-2} 
\end{figure}
To examine this trembling motion we use Eq. (\ref{eq:p17}) and solve it numerically for expectation  values of both $x$ and $y$ coordinates. 
  Figs.\ref {Fig-2} $(a),(c),(e),(g)$ and  Figs.\ref {Fig-2} $(b),(d),(f),(h)$ show the oscillations in the   wave packet's center along $x$ and $y$ directions respectively. Results are plotted for two values of initial momentum  $k_{0x}$ = 0.6 and 0.8 i.e $a_0=k_{0x} \hspace{0.5mm} \textit l$=$1.2$ and $1.6$, given by solid and dotted lines  for four different values of pseudospin polarization. It is  clear from  Fig.\ref {Fig-2} that an increase in momentum results in a decrease in amplitude of Zitterbewegung oscillations without any other change in the behaviour of ZB. In further discussion, ZB oscillations corresponding to different pseudospin polarizations will be referred to as $\bar x_{\{c_1,c_2\}}$ and $\bar y_{\{c_1,c_2\}}$.
\\
The ZB oscillations shown in Fig. \ref{Fig-2} are of the order of $nanometer$ (easily detectable), except $x(t)_{\{1,0\}}$ and $x(t)_{\{1,i\}}$ which are of the order of $0.1 \hspace{0.8mm} pm$. On comparing the ZB oscillations $x(t)_{\{1,0\}}$, $x(t)_{\{1,1\}}$, $x(t)_{\{1,i\}}$, we have concluded that when the direction of initial momentum ($x$-axis) is perpendicular to initial pseudospin polarizations ($z,y$-axis) then the ZB oscillations carry very small amplitude in the direction of initial momentum.  Note that, ZB Oscillations in Fig. \ref{Fig-2} $(g)$ and  Fig. \ref{Fig-2} $(h)$ are very similar to those of Fig. \ref{Fig-2} $(c)$ and  Fig. \ref{Fig-2} $(f)$ respectively. This is because  pseudospin polarization $\{ 1 , e^{i\frac{\pi}{4}}\} $ has projection on both $x$ and $y$ axes so ZB oscillations in this case have detectable amplitude in both directions. 
\\
In Figs.\ref{Fig-2} $(a)$-$(h)$ we have seen that  initially 
 the  amplitude of the Zitterbewegung oscillations  increases, then these oscillations seems to die out but  reappear for all pseudospin polarizations. For example in Fig.\ref{Fig-2} $(c)$ these oscillations reappear at $t\simeq18,55,84 \hspace{0.5mm}\tau_o ...$ . Hence when there is a  magnetic field applied to  the system, Zitterbewegung oscillations are not transient; rather they are recurrent. Also, in the presence of the magnetic field,  several ZB frequencies appear (see  Figs.\ref{Fig-2} $(a)$-$(h)$). 
This is different from the ZB phenomenon observed in monolayer graphene without magnetic field, in which ZB oscillations are transient having a single frequency.\cite{PhysRevB.78.235321}
\\
Hence, the results in hand are in agreement with the previous studies in both respects i.e., (1) in the presence of a magnetic field ZB oscillations are permanent and (2) ZB oscillations strongly depend on initial pseudospin polarization. \cite{rusinZBinB2008} \cite{PZB}
\begin{figure}[t]
\centering
\includegraphics[scale=0.41]{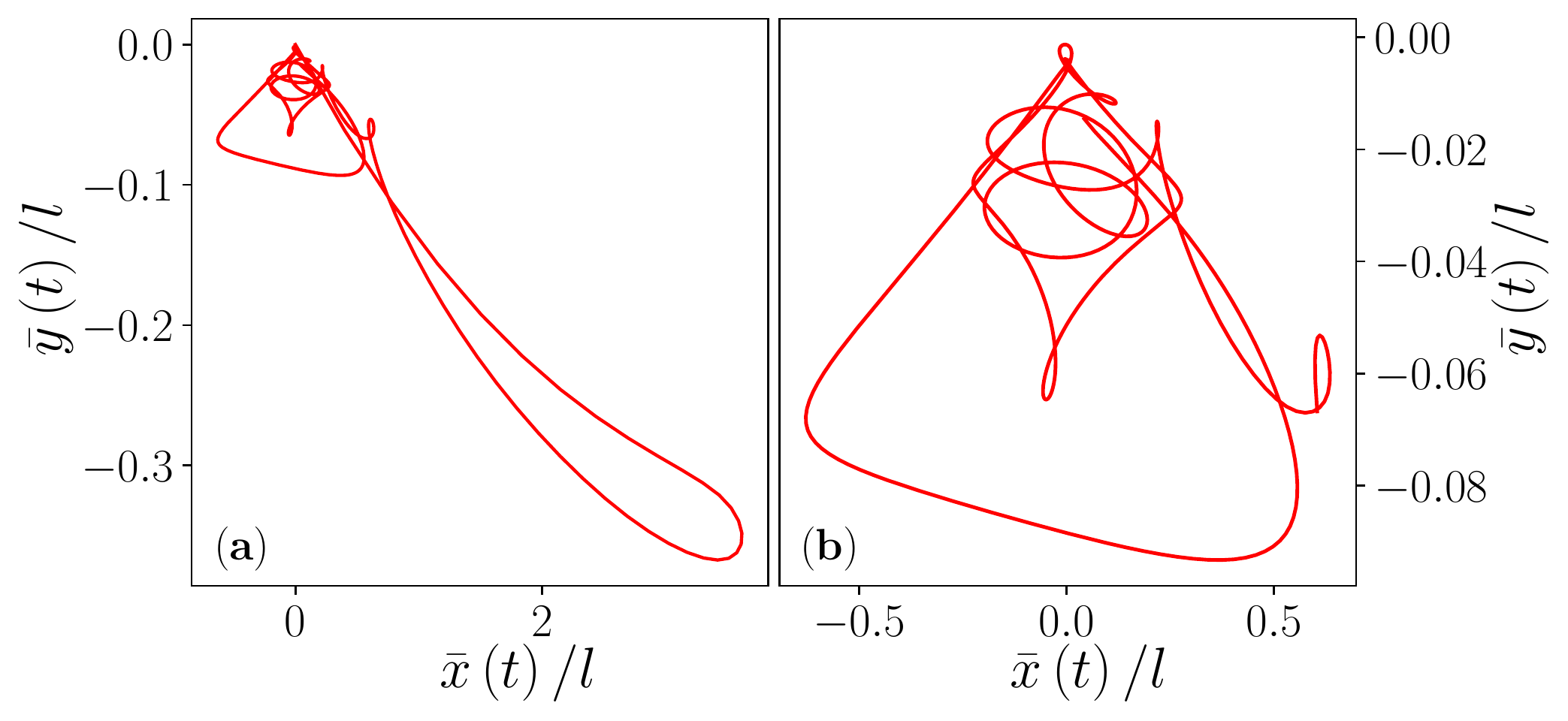}
\caption{Average coordinates $\bar x(t)$ versus $\bar y(t)$  corresponding to  pseudospin polarization $\{c_1,c_2\}$=$\{1,1\}$ for momentum $a_0=1.2$ with $B=3.3 \hspace{1mm} Tesla$  (a) ZB Trajectory for $t=0$ to $t =100 \tau_o (0.2\hspace{0.8mm} ps)$. (b) Zoomed view of the same trajectory for $t=7\tau_o$ to $t =100 \tau_o$  . }
\label{Fig-3} 
\end{figure}
Finally, in Fig. \ref{Fig-3} we have plotted the average coordinates   $\bar x(t)_{\{c_1,c_2\}}$ and $\bar y(t)_\{c_1,c_2\}$ against each other to study the ZB trajectory of the center of the wave packet corresponding to the Figs.\ref{Fig-2} $(c),(d)$. Initially due to large ZB, the center of the packet sweeps a large area. For a better understanding,  we have shown a zoomed view of the ZB trajectory  in Fig. \ref{Fig-3} $(b)$. Due to the presence of recurrent ZB,  these trajectories do not disappear with time (infinite trajectories). 
\\
It is well known that the electronic band structure of graphene can be modified  by introducing external ID potentials using nanopatterning. \cite{5391729} These 1D potentials can not only modify the energy spectrum of the system but they also help us to control the charge transport properties of the system. \cite{PhysRevLett.121.207401, RevModPhys.81.109, nl801752r, PhysRevLett.113.026802, Park2008AnisotropicBO}. Potentials such as
antidot lattices can be carved on graphene by various techniques  and lattice parameters can be tuned.\cite{PhysRevLett.100.136804, PhysRevB.86.045445} With the aim of controlling electron propagation in graphene along a 1D channel we introduce a 1D antidot lattice in the following section.
\section{\label{sec:level2}Graphene Antidot lattice in the presence of a magnetic field: Landau Minibands}
\begin{figure}[t]
\centering
\includegraphics[scale=0.27]{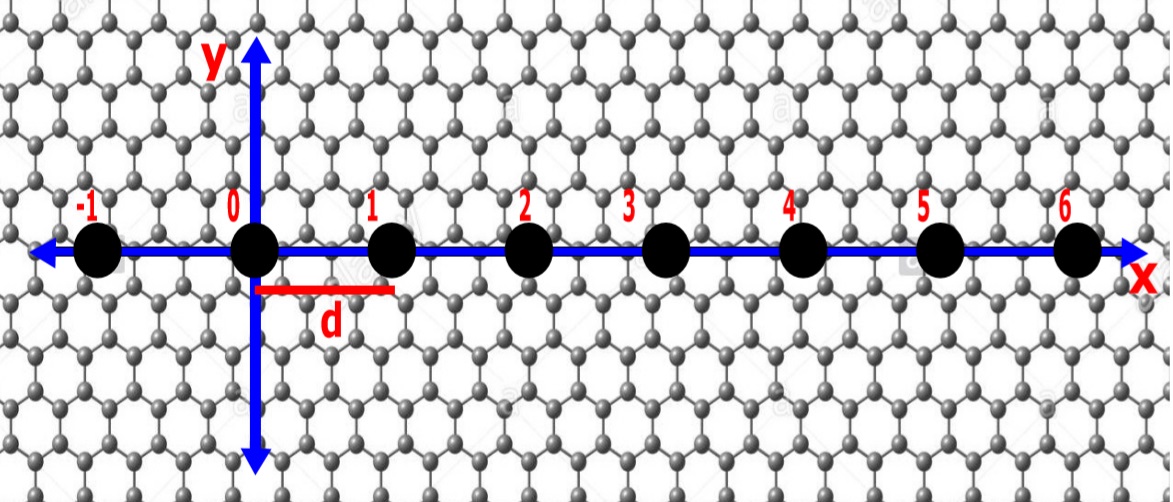}
\caption{Two dimensional graphene sheet having a one dimensional lattice of quantum antidot barriers at $y=0$. Solid black circles represent periodically placed quantum antidots along the $x$-axis.  $ a^2$ is the dot area and $d$ is the uniform separation  of  the periodically placed quantum antidot potential barriers.} 
\label{Fig-4} 
\end{figure}

\begin{figure}[!t]
\centering
\includegraphics[scale=0.54]{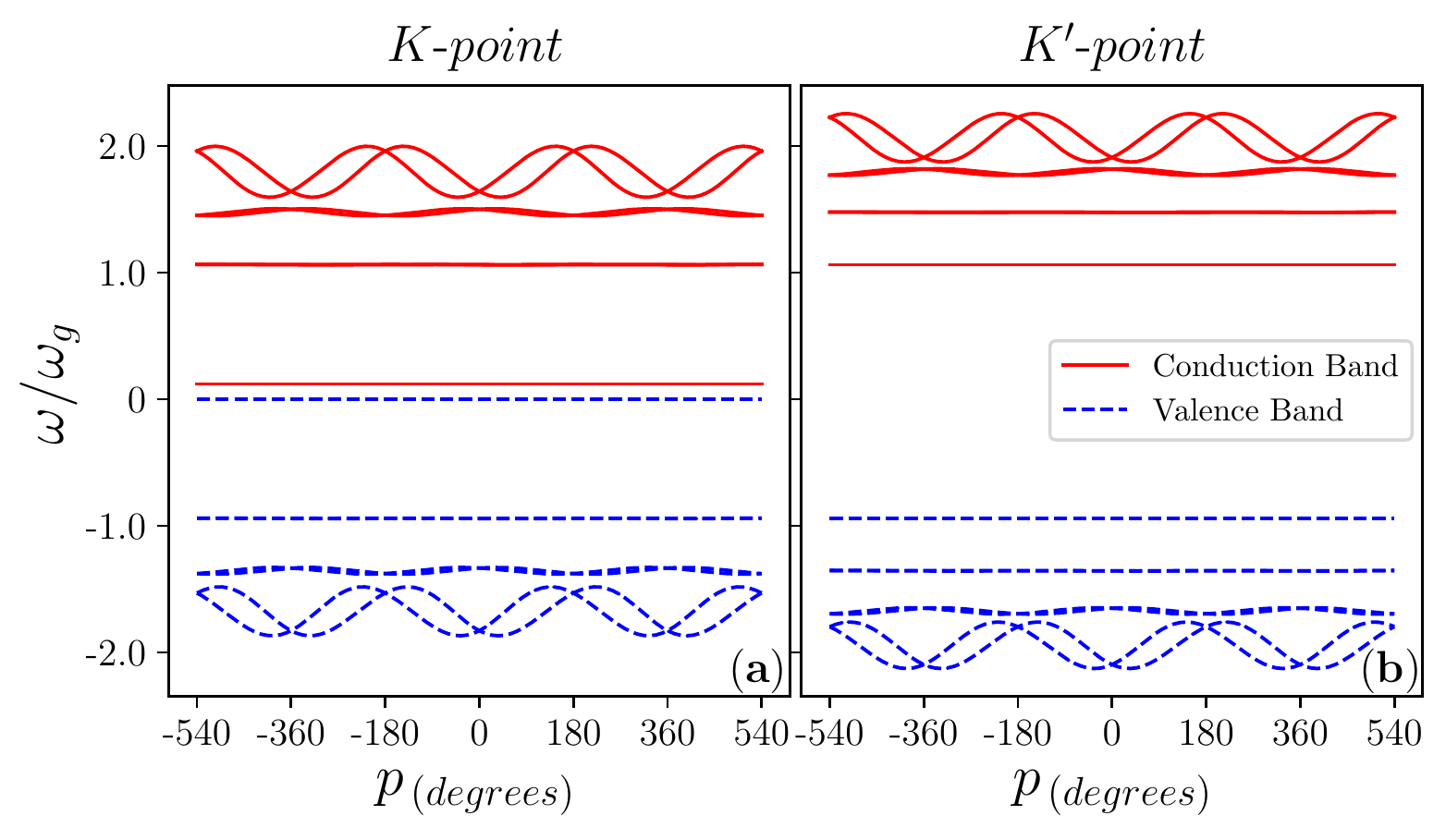}  
\caption{Landau minibands ($n=0-3$) in Landau quantized graphene having a one dimensional antidot lattice with  antidot potential $U_0$=$100 \hspace{0.8mm} meV$, antidot radius $a$=$10 \hspace{0.8mm} nm$ and spacing $d$=$100 \hspace{0.8mm} nm$ in the presence of magnetic field $B$=$3.3 \hspace{1mm} Tesla$. (a) $K$ point. (b) $K^\prime$ point. }
\label{Fig-5} 
\end{figure}

We consider a two-dimensional graphene sheet having a one-dimensional lattice of quantum antidot potential barriers, with a  quantizing magnetic field B, which is perpendicular to the plane of the graphene sheet.
\\
We model the   antidot array as  a row of Dirac delta functions.
Following the Kronig-Penny model for a quantum antidot lattice, we introduce an infinite array of  identical  quantum antidot potential barriers periodically spaced along the $x$-axis at
$x_{n^\prime}=n^\prime d , y\equiv0 $ 
as
\begin{equation} \label{p23}
U(\textbf r)=U(x,y)=\al\sum_{n^\prime=-\infty}^\infty \delta(x-n^\prime  d)\delta(y) \hspace{1mm} ,
\end{equation}
where $\al>0$ , $n^\prime =-\infty$ to $+\infty$ and
\begin{equation} \label{p24}
\al=U_0 \hspace{1mm} a^2
\end{equation}
where $U_0$ is the barrier height, $ a^2$ is area  and $d$ is the uniform separation  of  the periodic quantum antidot potential barriers as shown in Fig.\ref{Fig-4}
\\
The Green's function $\mathcal G(x_1;x_2;0,0;\omega)$ for the two dimensional Kronig-Penney-like model having a one dimensional antidot lattice, for an electron propagating directly along  the axis of the antidot lattice ($y\equiv 0\equiv y_1\equiv y_2$ and suppressing further reference to $y$) is given by \cite{Minibandss}
\begin{multline} \label{p25}
\mathcal G(x_1;x_2;\omega) = \mathcal G^0(x_1;x_2;\omega) +  \frac{\al d}{2\pi} \sum_{n^\prime =-\infty}^\infty \mathcal G^0(x_1;n^\prime  d;\omega)
\\
\times  \int_{-\frac{\pi}{d}}^{\frac{\pi}{d}} dp\hspace{2mm} e^{-\textit ipn^\prime d}\left[ 1-\al \hspace{1mm} \dot{\tilde{\mathcal G}}^0(p;0,0;\omega)\right]^{-1} \tilde{\mathcal G}^0(p;x_2;\omega).
\end{multline}
Here, $\mathcal G^0(\textbf r_1,\textbf r_1^\prime)$ is the Green's function for  graphene  in a perpendicular magnetic field in the complete absence of quantum antidot potential barriers, which is given by Eqs. (\ref{eq:p2}),(\ref{eq:p5}) and (\ref{eq:p6}), while the overhead dot represents the  spatially translationally invariant Green's function i.e. $\mathcal G^0(m^\prime  d,n^\prime  d;\omega) = \dot{\mathcal G}^0([m^\prime  -n^\prime ]d;\omega)$ where $m^\prime$ and $n^\prime$ are  integers. Also the Green's function $\tilde{\mathcal G}(p)$ can be expanded    in a Fourier series due to the periodicity of the lattice, given by \cite{Minibandss}
\begin{equation}\label{p26}
\tilde{\mathcal G}(p) = \sum_{r=-\infty}^\infty e^{\textit iprd} \mathcal G(rd)
\end{equation}
where $r=0,1,2,3$... are integers, with
\begin{equation}\label{p27}
\mathcal G(m^\prime d) = \frac{d}{2\pi} \int_{-\frac{\pi}{d}}^{\frac{\pi}{d}} dp\hspace{1mm} e^{-\textit ipdm^\prime } \tilde{\mathcal G}(p),
\end{equation}
and
\begin{equation}\label{p29}
\dot{\mathcal G}^0([m^\prime -n^\prime ]d) = \frac{d}{2\pi} \int_{-\frac{\pi}{d}}^{\frac{\pi}{d}} dp \hspace{1mm} e^{-\textit ip\left(m^\prime -n^\prime \right)d} \hspace{2mm} \dot{\tilde{\mathcal G}}^0(p).
\end{equation}
 It is important to note that the Peierls phase factor is $C(\textbf r, \textbf r^\prime)$=$1$ for the case  involving propagation directly  along the axis of the antidot lattice  (our choice $y \equiv 0\equiv y_1\equiv y_2$  results in $\textbf r \parallel \textbf r^\prime$). Therefore  $C(\textbf r, \textbf r^\prime)$  does not appear in  Eq. (\ref{p25}) and the eigen-energy spectrum given by poles of Eq. (\ref{p25}) is unaffected by  $C(\textbf r, \textbf r^\prime)$.
\\
The energy spectrum of this system  can be obtained from vanishing of the frequency poles of the Green's function of Eq. (\ref{p25}):
\begin{equation} \label{p30}
Det( I_2-\al \hspace{1mm} \dot{\tilde{\mathcal G}}^0(p;0,0;\omega)) = 0.
\end{equation}
Eqs. (\ref{eq:p2}) and (\ref{eq:p5}) taken jointly with  Eq. (\ref{p26}) yields  ($Y=0$, $X^2+Y^2 = X^2 = (rd)^2$)

\begin{multline}\label{eq:2.31a}
\dot{\tilde{\mathcal G}}^0_{11\atop 22}(p;0,0;\omega)_K= \frac{1}{4\pi\hbar\ga^2}\hspace{0.8mm} \omega \hspace{0.8mm} \sum_{r=-\infty}^\infty e^{i prd}  \hspace{0.8mm}   e^{-\frac{\omega^2_g}{8\ga^2}\left(rd \right)^2}
\\
\times \sum_{n=0}^\infty
 \frac{ L_n\left[\frac{\omega^2_g}{4\ga^2}\left(rd \right)^2\right]}{\frac{\omega^2}{\omega^2_g}-n}
\end{multline}
and
\begin{multline}\label{eq:2.32b}
\dot{\tilde{\mathcal G}}^0_{21\atop 12}(p;0,0;\omega)_K= \frac{\textit i  \hspace{0.6mm} \omega_g^2}{8\pi\hbar\ga^3} \hspace{0.8mm} \sum_{r=-\infty}^\infty  \left(rd \right) \hspace{0.8mm}  e^{i prd}   \hspace{1mm}  e^{-\frac{\omega^2_g}{8\ga^2}\left(rd\right)^2} 
\\
\times \sum_{n=1}^\infty
 \frac{ L_{n-1}^1 \left[\frac{\omega^2_g}{4\ga^2}\left(rd \right)^2\right]}{\frac{\omega^2}{\omega^2_g}-n}.
\end{multline} 
for the $K$ point in graphene. It is clear from Eqs. (\ref{eq:2.31a}) and  (\ref{eq:2.32b}) that $\dot{\tilde{\mathcal G}}^0_{11}(p;0,0;\omega)_K=\dot{\tilde{\mathcal G}}^0_{22}(p;0,0;\omega)_K$ and $\dot{\tilde{\mathcal G}}^0_{12}(p;0,0;\omega)_K=\dot{\tilde{\mathcal G}}^0_{21}(p;0,0;\omega)_K$. Hence Eq. (\ref{p30}) can be written as
\begin{equation} \label{p30a}
1-2\al \dot{\tilde{\mathcal G}}^0_{11}+\al^2 \left({\dot{\tilde{\mathcal G}}^0_{11}}^2 - {\dot{\tilde{\mathcal G}}^0_{12}}^2\right) = 0.
\end{equation}
Note that since the  antidot radius is very small (i.e $\frac{\al \omega_g}{4\pi\hbar \ga^2}\ll 1$), 
a root of Eq.   (\ref{p30a}) $\omega$ approaches the pole position i.e $\omega \rightarrow \omega_n$, so that the $n$-$th$ pole has the primary influence in determining the eigen-energy root $\omega_n$. Therefore, we can make a reasonable first approximation by dropping all other terms of the $n$-sum. Also,  for $\frac{ \omega_g d}{8\ga}>1$ , it suffices to keep only $r=-1,0,1$ terms of the $r$-sum in Eqs.  (\ref{eq:2.31a})  and (\ref{eq:2.32b}). This imposes following condition on antidot spacing
\begin{equation}  \label{p35}
d  > \frac{145}{\sqrt{B \hspace{1mm}(Tesla)}} \hspace{1mm}(nm).
\end{equation}
Finally, Eq. (\ref{p30a}) can be written as
\begin{equation}  \label{p33}
1-2 \frac{\omega_g\hspace{0.6mm}\Omega\hspace{0.6mm}\omega_n}{\omega_n^2-n\omega_g^2} +   \frac{\omega_g^2\hspace{0.6mm}\Omega^2\hspace{0.6mm}\omega_n^2}{\left(\omega_n^2-n\omega_g^2\right)^2}
-  \frac{\omega_g^4\hspace{0.6mm}\kappa^2}{\left(\omega_n^2-n\omega_g^2\right)^2}
\end{equation}
where we have defined
\begin{equation}  \label{eq:2.25}
\Omega = \frac{\al \hspace{0.8mm} \omega_g }{4\pi\hbar\ga^2}\hspace{0.8mm} \left[1+ 2 \cos{pd} \hspace{1.2mm}   e^{-\frac{\omega^2_g}{8\ga^2}d^2}   L_n\left(\frac{\omega^2_g}{4\ga^2}d^2 \right)\right]
\end{equation}
and
\begin{equation}  \label{eq:2.25a}
\kappa = \frac{\al \hspace{0.8mm} \omega_g }{4\pi\hbar\ga^2}\hspace{0.8mm} \left( \frac{\omega_g\hspace{0.8mm}d}{\ga}\right) \sin{pd} \hspace{1.2mm}   e^{-\frac{\omega^2_g}{8\ga^2}d^2}   L_n\left(\frac{\omega^2_g}{4\ga^2}d^2 \right).
\end{equation}
The four roots of Eq. (\ref{p33}) describe the energy spectrum at the $K$ point of graphene having a 1-D antidot lattice placed in a uniform normal magnetic field: 
\begin{equation}  \label{eq:2.24}
\omega_{n,K} = \frac{\Omega \pm \sqrt{\Omega^2 + 4(n\pm^\prime\kappa)}}{2} \hspace{0.8mm} \omega_g .
\end{equation}
Similarly, for $K^\prime$ point, $n$ will be replaced by $n+1$ on the right hand side of Eq. (\ref{eq:2.24}).
\\
In this, we have  the energy spectrum composed of broadened Landau levels (Landau minibands) for a graphene  antidot lattice in a quantizing magnetic field. Each Landau level has split into two branches and each branch  has broadened into a small continuous band (subband) of energy instead of a single energy. Fig. \ref {Fig-5} (a) and  Fig. \ref {Fig-5} (b) show the Landau minibands at the location of $K$ and $K^\prime$ points. The broadening is so small that it can not be observed with the naked eye, so  we  multiplied a broadening factor $\beta=200$ with the oscillatory terms $\cos (pd)$ and $\sin (pd)$.  The  parameter $\beta$ introduces an increase in the amplitude of the minibands to facilitate  observation of the broadening of the Landau minibands;  i.e we used
\begin{equation*} 
\Omega = \frac{\al \hspace{0.8mm} \omega^2_g }{4\pi\hbar\ga^2}\hspace{0.8mm} \left(1+ 2\hspace{0.5mm} \beta  \hspace{0.8mm}  \cos{pd} \hspace{1.2mm}   e^{-\frac{\omega^2_g}{8\ga^2}d^2}   L_n\left[\frac{\omega^2_g}{4\ga^2}d^2 \right]\right)
\end{equation*}
and 
\begin{equation*} 
\kappa = \frac{\al \hspace{0.8mm} \omega_g }{4\pi\hbar\ga^2}\hspace{0.8mm} \left( \frac{\omega_g\hspace{0.8mm}d}{\ga}\right) \beta\hspace{0.6mm} \sin{pd} \hspace{1.2mm}   e^{-\frac{\omega^2_g}{8\ga^2}d^2}   L_n\left(\frac{\omega^2_g}{4\ga^2}d^2 \right)
\end{equation*}
to exhibit the Landau minibands plotted in Fig. \ref {Fig-5}.
The  energy spectra of the $K$ and $K^\prime$ points  differs by a unit shift even in the presence of an antidot lattice. The presence of antidots induced a gap between conduction and valence bands at both $K$ and $K^\prime$ points. In the case of Fig. \ref {Fig-5}, if we consider $\beta=1$ with all other parameters having same values, then the gap opened at the location of $K$ and $K^\prime$ points is $7.96\hspace{0.8mm} meV$ and $132\hspace{0.8mm} meV$ respectively. This gap between conduction and valence bands increases with the increase in strength of magnetic field and reaches $36.25\hspace{0.8mm} meV$ and $283.4\hspace{0.8mm} meV$ for $K$ and $K^\prime$ points respectively when $B$=$15 \hspace{1mm} Tesla$. Similarly this gap shows an increase with increase in strength of the antidot lattice.
\\
Note that when the antidot strength $\al$ approaches  zero (which means no antidot lattice), $\Omega$ and $\kappa$ also approaches  zero, i.e
\begin{equation*}
 \lim_{\al\rightarrow 0} \Omega = 0 =  \lim_{\al\rightarrow 0} \kappa .
\end{equation*}
In this case  Eq. (\ref{eq:2.24}) reduces to the case of a discrete eigen-energy spectrum of graphene  in a normal, uniform magnetic field. The width of Landau minibands for the $K$ point  of graphene is
\begin{equation*}
\Delta\omega_{n,K} = [\omega_{n,K}]_{p=\frac{2\pi}{d}} -[\omega_{n,K}]_{p=\frac{\pi}{d}}.
\end{equation*}
Evaluation of this expression for the width of Landau minibands yields 
\begin{multline}  \label{eq:2.27}
\frac{\Delta\omega_{n,K}}{\omega_g} =   \al^\prime \hspace{0.8mm} \varrho   \pm \frac{1}{2} \sqrt{ \al^{\prime 2}(1+\varrho)^2+ (2\sqrt{n})^2 }
\\
\mp \frac{1}{2} \sqrt{ \al^{\prime 2}(1-\varrho)^2+ (2\sqrt{n})^2} \hspace{1mm} ,
\end{multline}
where we have defined 
\begin{equation*}   \label{eq:2.28}
\al^\prime = \frac{\al\hspace{0.8mm}\omega_g}{4 \pi \hbar  \ga^2}      \qquad           \varrho =2 \hspace{0.6mm}e^{-\frac{\omega^2_g}{8\ga^2}d^2} \hspace{1.3 mm}  L_n\left[\frac{\omega^2_g}{4\ga^2}d^2 \right]  .
\end{equation*}
Upper and lower signs  are used for conduction and valence bands respectively. For the  width of the Landau minibands for the $K^\prime$ point, we have $ (2\sqrt{n+1})^2$ in place of $(2\sqrt{n})^2$ in Eq. (\ref{eq:2.27}).
\\
To investigate electron propagation in the system under study, first we have to find the Green's function matrix of the system in time representation. This is evaluated in section V.
\section{\label{sec:level2} Green's Functions: Frequency and Time Representation}
 To evaluate the full Green's  function matrix $ \mathcal G(x_1;x_2;\omega)_K$ given in Eq. (\ref{p25}), we have to find the
$ \mathcal G^0(x_1;x_2;\omega)$,
$ \mathcal G^0(x_1;nd;\omega)$,
 $\dot{\tilde{\mathcal G}}^0(p;0,0;\omega)$  and 
$ \tilde{\mathcal G}^0(p;x_2;\omega)$ matrices. 
These four matrix Green's functions  can be easily determined using $ \mathcal G^0(x_1;x_2;\omega)$, which is the Green's function in the presence of perpendicular and uniform magnetic field in absence of an antidot lattice given in  Eqs. (\ref{eq:p2}) and (\ref{eq:p5}).
\\
For  propagation along the antidot lattice i.e the $x$ axis only ($y_1$=0=$y_2$, $X$=$x_1-x_2 $ and $C(\textbf r_1 ;\textbf r_2)$=$1$ ),  Eqs. (\ref{eq:p2}) and (\ref{eq:p5}) reduce to (for $K$-point )
\begin{multline}\label{eq:2.29}
\mathcal G^0_{ 11\atop 22}(x_1,x_2;\omega)_K= \frac{1}{4\pi\hbar\ga^2}\hspace{0.8mm} \omega \hspace{0.8mm} e^{-\frac{\omega^2_g}{8\ga^2}\left(x_1-x_2 \right)^2}
\\
\times \sum_{n=0}^\infty
 \frac{ L_n\left[\frac{\omega^2_g}{4\ga^2}\left(x_1-x_2 \right)^2\right]}{\frac{\omega^2}{\omega^2_g}-n}
\end{multline}
and
\begin{multline}\label{eq:2.30}
 \mathcal G^0_{21\atop 12}(x_1,x_2;\omega)_K= \frac{\textit i  \hspace{0.6mm} \omega_g^2}{8\pi\hbar\ga^3} \hspace{0.8mm} e^{-\frac{\omega^2_g}{8\ga^2}\left(x_1-x_2 \right)^2}  \left(x_1-x_2 \right)
\\
\times \sum_{n=1}^\infty
 \frac{ L_{n-1}^1 \left[\frac{\omega^2_g}{4\ga^2}\left(x_1-x_2 \right)^2\right]}{\frac{\omega^2}{\omega^2_g}-n}
\end{multline} 
respectively. Matrix elements of $ \mathcal G^0(x_1;n^\prime  d;\omega)_K$ can be obtained by taking $x_2$=$n^\prime  d$ in Eqs. (\ref{eq:2.29}) and  (\ref{eq:2.30}), while matrix elements  of $\dot{\tilde{\mathcal G}}^0(p;0,0;\omega)_K$ are  given in Eqs.   (\ref{eq:2.31a})  and (\ref{eq:2.32b}).
\\
Similarly, the matrix  $ \tilde{\mathcal G}^0(p;x_2;\omega)_K$ can be obtained by using $x_1$=$rd$  in Eqs. (\ref{eq:2.29}) and  (\ref{eq:2.30}) jointly with Eq. (\ref{p26}).
Note that for  $\frac{ \omega_g d}{8\ga}>1$ , it suffices to keep only $r=-1,0,1$  of the $r$-sum in the expressions for matrix elements of  $\dot{\tilde{\mathcal G}}^0(p;0,0;\omega)_K$ and $ \tilde{\mathcal G}^0(p;x_2;\omega)_K$,  as discussed earlier. The above four matrix Green's functions completely determine the full Green's function   $ \mathcal G(x_1;x_2;\omega)_K$ given in Eq. (\ref{p25}). 
\\
As discussed earlier,  for temporal evolution of the wave packet we require the time representation of the Green's function. Hence, to find the time representation of the full Green's function, we have to  take the Fourier  transform of  $ \mathcal G(x_1;x_2;\omega)_K$ matrix  using Eqs.  (\ref{eq:p8}) and (\ref{p25}):
\begin{multline} \label{eq:2.38}
\mathcal G_{\mu\nu}(x_1,x_2; t)_{K}  =  \int_{-\infty}^\infty  d\omega \hspace{0.7mm} e^{-\textit i \omega t}  \mathcal G^0(x_1;x_2;\omega) 
\\
+  \frac{\al d}{2\pi}  \int_{-\infty}^\infty  d\omega \hspace{0.7mm}
 e^{-\textit i \omega t}  \sum_{n^\prime  =-\infty}^\infty \mathcal G^0(x_1;n^\prime  d;\omega)
\\
\times  \int_{-\frac{\pi}{d}}^{\frac{\pi}{d}} dp\hspace{2mm} e^{-\textit ipn^\prime  d}\left[ 1-\al \hspace{1mm} \dot{\tilde{\mathcal G}}^0(p;0,0;\omega)\right]^{-1} \tilde{\mathcal G}^0(p;x_2;\omega).
\end{multline}
 There are two integrals in the above equation. The first integral was evaluated in   Eqs. (\ref{eq:p9}) and (\ref{eq:p10}) as
\begin{multline}  \label{eq:2.39}
\mathcal G^0_{11 \atop 22}\left( x_1,x_2;t\right)=-i \eta_+(t) \frac{ \omega_g^2}{4 \hspace{0.5mm}\pi \hspace{0.5mm}\gamma^2} \hspace{1.2 mm}  e^{-\frac{\omega^2_g}{8\ga^2}\left(x_1-x_2 \right)^2} \hspace{1.2 mm}
\\
\times \sum_{n=0}^\infty
 L_n\left[\frac{\omega^2_g}{4\ga^2}\left(x_1-x_2 \right)^2\right]  \hspace{1mm}
\cos \left(\frac{\omega_g \hspace{0.8mm} l}{\ga} \hspace{0.5mm} \frac{t}{\tau_o} \hspace{0.4mm} \sqrt n \hspace{0.8mm}\right) 
\end{multline}
and
\begin{multline}  \label{eq:2.40}
\mathcal G^0_{21 \atop 12}\left( x_1,x_2;t\right)=-i \eta_+(t) \frac{ \omega_g^3 \hspace{0.5mm} l}{8 \hspace{0.2mm}\pi \hspace{0.2mm}\gamma^3} \hspace{0.3 mm}   e^{-\frac{\omega^2_g}{8\ga^2}\left(x_1-x_2 \right)^2} \left(\frac{x_1-x_2}{\textit l}\right)
\\
\times \sum_{n=1}^\infty \frac{ L_{n-1}^1\left[\frac{\omega^2_g}{4\ga^2}\left(x_1-x_2 \right)^2 \right] }{\sqrt n}
\ \sin \left(\frac{\omega_g \hspace{0.8mm} l}{\ga} \hspace{0.5mm} \frac{t}{\tau_o} \hspace{0.4mm} \sqrt n \hspace{0.8mm}\right) .
\end{multline}
Evaluation of the second term of Eq. (\ref{eq:2.38}) is a lengthy process. The matrix elements $V_{i,j}(t)$ ($i.j$=$1,2$) of  the second term in time representation are given by (see Appendix A)
\begin{multline} \label{eq:2.42}
 V_{11 \atop 22} = -i \eta_+(t) \frac{ \omega_g}{2\pi} \sum_{n^\prime  =-\infty}^\infty \Bigg\{  \ga_1 \left[c_{11 \atop 22}(\eta)\right]_{\eta=0} +   \sum_{n=1}^\infty  
\\
 \times  \cos \left(\frac{\omega_g \hspace{0.8mm} l}{\ga} \hspace{0.5mm} \frac{t}{\tau_o} \hspace{0.4mm} \sqrt n \hspace{0.8mm}\right)
  \Bigg[ \ga_1 L_n \left[\Upsilon \right] c_{11 \atop 22}(\eta) 
\\
 +  \frac{\ga_2}{\sqrt{n}}          L_{n-1}^1 \left[\Upsilon\right] c_{21 \atop 12} (\eta) \Bigg]_{\eta=\sqrt{n}} \Bigg\}
\end{multline}
and 
\begin{multline} \label{eq:2.43}
 V_{12 \atop 21} = - \eta_+(t) \frac{ \omega_g}{2\pi} \sum_{n^\prime  =-\infty}^\infty  \Bigg\{i \ga_1 \left[c_{12 \atop 21}(\eta)\right]_{\eta=0} +   \sum_{n=1}^\infty  
\\  \times \sin \left(\frac{\omega_g \hspace{0.8mm} l}{\ga} \hspace{0.5mm} \frac{t}{\tau_o} \hspace{0.4mm} \sqrt n \hspace{0.8mm}\right) 
  \Bigg[ \ga_1 L_n \left[\Upsilon\right] c_{12 \atop 21}(\eta) 
\\
+   \frac{\ga_2}{\sqrt{n}}          L_{n-1}^1 \left[\Upsilon\right] c_{22 \atop 11} (\eta) \Bigg]_{\eta=\sqrt{n}} \Bigg\}
\end{multline}
where 
\begin{equation} \label{eq:2.44}
\Upsilon= \frac{\omega^2_g l^2}{4\ga^2}\left(\frac{x_1-n^\prime d}{l} \right)^2,
\qquad
\ga_1= \frac{\al \omega_g}{4\pi\hbar\ga^2}\hspace{0.8mm} \hspace{0.8mm}   \hspace{0.8mm}   e^{-\frac{\Upsilon}{2}},
\end{equation}
\begin{equation} \label{eq:2.45}		
\ga_2=\frac{\textit i  \hspace{0.6mm} \al \omega_g^2 l}{8\pi\hbar\ga^3} \hspace{0.8mm}  \left(\frac{x_1-n^\prime d}{l} \right) \hspace{0.8mm}  \hspace{1mm}  e^{-\frac{\Upsilon}{2}},
\end{equation}
and $c_{ij}(\eta)$ (where $\eta$=$\frac{\omega}{\omega_g}$) are matrix elements of (see Eq. (\ref{C8}): notation $q\equiv pd$)
\begin{equation*}
\int_{-\pi}^{\pi} dq \hspace{1mm}  e^{{-\textit iqn^\prime }} \left[ I-\al \hspace{1mm} \dot{\tilde{\mathcal G}}^0(p;0,0;\eta)\right]^{-1} \tilde{\mathcal G}^0(p;x_2;\eta)
\end{equation*}
These considerations yield the  time representation   of the full Green's function, i.e. $\mathcal G(x_1,x_2; t)_{K}$  for a   Landau-quantized monolayer graphene having a one dimensional antidot lattice.
In  the section VI, this Green's function $\mathcal G(x_1,x_2; t)_{K}$ will be employed to study the  temporal dynamics of a wave packet in the lattice system.
\section{\label{sec:level2} Wave packet dynamics along A one dimensional antidot lattice}
\begin{figure}[!t]
\centering
\includegraphics[scale=0.68]{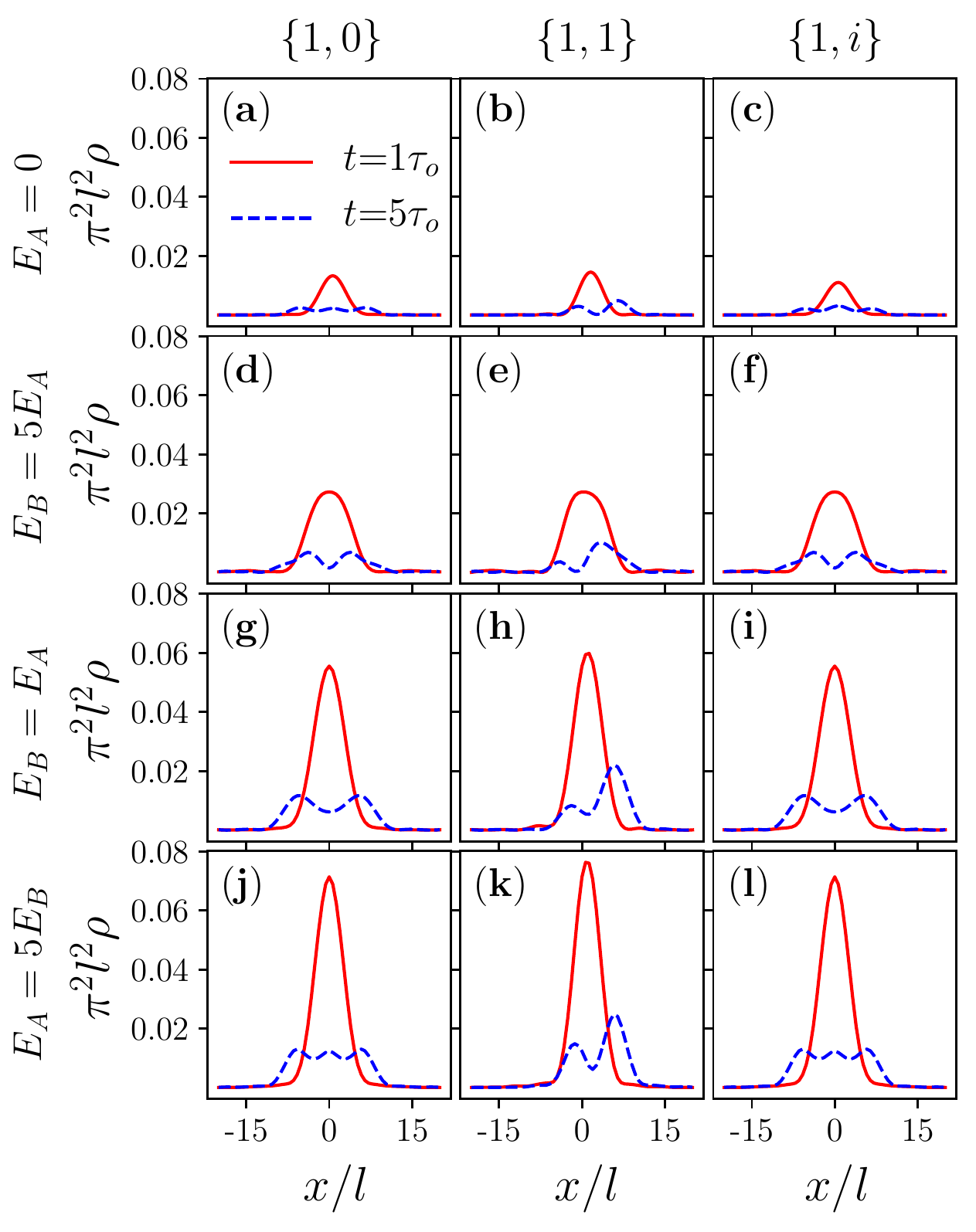}  
\caption{Electron probability density $\rho(x,0,t)$ for $K$ point along the $x$ axis  in the presence of a uniform magnetic field $B=3.3 \hspace{0.8mm} Tesla$ having strength $E_B=66 \hspace{0.8mm} meV$ for $a_0=k_{0x}l=1.2$ with  $l=2 \hspace{0.8mm} nm$; also  antidot lattice parameters, spacing $d=100 \hspace{0.8mm} nm $ and radius $a=10 \hspace{0.8mm} nm $. Left to right: Three columns with initial pseudospin polarization $\{1,0\},\{1,1\}$ and $\{1,i\}$ respectively. Up to down: Four rows with antidot strength $E_A=U_0=0, 13, 66$ and $330 \hspace{0.8mm} meV$ respectively.  }
\label{Fig-6} 
\end{figure}

With the solution of Eq.  (\ref{eq:2.38}) in hand, in the form of Eqs.  (\ref{eq:2.39}), (\ref{eq:2.40}), (\ref{eq:2.42}) and (\ref{eq:2.43}), the temporal study of an electron wave packet propagating along the axis of the antidot lattice
 given by 
\begin{equation} \label{eq:p13a}
\psi(\textbf r,0) =  \frac{f(\textbf r) }{\sqrt{\mid c_1\mid ^2+\mid  c_2 \mid ^2}} 
\begin{pmatrix} 
c_1  \\
c_2  
\end{pmatrix}
\end{equation}
\begin{equation*} \label{eq:p14a}
f(\textbf r) = \frac{1}{l  \hspace{0.4mm} \sqrt\pi }  \hspace{0.5 mm}  exp\left(-\frac{x^2+y^2}{2\hspace{0.4mm} l^2}+\textit i k_{0x} x\right) \delta(y)
\end{equation*}
can be made using   Eqs. (\ref{eq:p15}) and (\ref{eq:p16}). In Fig. (\ref{Fig-6}), results for the probability density $\rho(x,0,t)$ of an electron along the axis of the antidot lattice $(x,y=0)$  are plotted for four different cases to examine the effect of the antidot lattice. 
\\
In Fig. \ref{Fig-6}, $\rho(x,0,t)$ is represented at times $t=1\tau_o$ and $t=5\tau_o$ with solid and dashed lines, respectively. Moving from left to right, the three columns correspond to three different pseudospin polarizations $\{c_1,c_2\}$, while from up to down the four rows represent the increase in antidot strength $E_A=U_0$ from zero $meV$ to $330 \hspace{0.8mm}meV$ with constant magnetic field strength $E_B=66 \hspace{0.8mm} meV$ ($E_B\equiv \hbar\omega_g$) at  $B=3.3 \hspace{0.8mm} Tesla$.
All three columns represent  the propagation of the wave packet along the $x$ axis ($y=0$). One can see that with the increase in antidot strength, the probability density of finding the electron starts increasing along the axis of antidot lattice. Row 1 corresponds to the situation when there is no antidot lattice and the wave packet propagates in graphene under the effect of the perpendicular magnetic field only. Similarly, rows 2, 3 and 4 correspond to $E_A=U_0= 13.2, 66$ and $330 \hspace{0.8mm} meV$ respectively. Clearly, the probability of finding an electron along the antidot direction increases with the introduction of the antidot lattice on a Landau quantized graphene sheet; this can be treated as the propagation of a wave packet through a quantum antidot wire, which is clearly supporting the propagation through it. Also, the probability density gets more confined with increase of antidot strength; this can be seen in rows 2, 3 and 4 of Fig. \ref{Fig-6}. Moving from up to down across the rows 2 to 4, the spread of the packet gets smaller with a clear increase in magnitude of probability density, which means that the probability of finding an electronic current along the antidot increases. This may be referred to as collimation of the electronic beam along the axis of antidot lattice.
\\
Further, we explore the effect of  initial pseudospin polarization on the wave packet dynamics in the presence of an antidot lattice, as shown in  Fig. \ref{Fig-6}. In Fig. \ref{Fig-6}, it can be seen that propagation of the wave packet is strongly affected by the change of pseudospin polarization across the rows. Columns 1 and 3 correspond to pseudospin $\{1,0\}$ and $\{1,i\}$ respectively and the propagation corresponding to these two columns is  similar because the  pseudospins are both perpendicular to the axis of the antidot lattice, and the center of the wave packet does not propagate with an increase in time; it is always located at $x=0$ without being affected by antidot strength, as shown in Figs. \ref{Fig-6} $(a),(d),(g),(h)$ and Figs. \ref{Fig-6} $(c),(f),(i),(l)$. But, one can see in Figs. \ref{Fig-6} $(b),(e),(h),(k)$ (column 2) that the center of the wave packet propagates along the axis of the antidot lattice when the initial pseudospin polarization is along the axis of the antidot lattice i.e. $x$ axis, at $t=5\tau_o$ the dashed lines clearly show the propagation of the wave packet when compared with the solid lines at $t=1\tau_o$. Hence, the wave packet only propagates along the one dimensional antidot lattice when the initial pseudospin polarization is parallel to the axis of antidot lattice.
\\
\textbf{Experimental relevance:}  We now address the question of experimental relevance of this work. In this regard, we note that it is possible to initialize the pseudospin and study  the dynamics of an electron wave packet in graphene by means of pump-probe laser spectroscopy. Experiments have been performed where pseudospin initialization and its subsequent relaxation have been probed on the femtosecond scale in graphene. \cite{PhysRevB.95.241412, PhysRevB.96.020301, PhysRevB.92.165429} Recently it has been shown that pesudospin can be manipulated by a coupled waveguide-cavity configuration in graphene. \cite{PhysRevResearch.2.033406}
\\
Another possibility is the creation of artificial graphene (honeycomb lattices) in cold atom systems. In these systems, it is possible to generate effective fields and study transport in graphene-like structures \cite{articlenat11, articlegomes, PhysRevLett.104.063901, articlemulti}
\\
Recently it has become possible to treat pseudospin as a real measurable  angular momentum at par with electron spin in photonic graphene (honeycomb array of evanescently coupled waveguides). \cite{PS2, articlegomes} These photonic systems allow a great degree of control of initial conditions in the study of wave packet dynamics.
\section[h!]{\label{sec:level2}Conclusions}
We have studied the evolution of a two dimensional Gaussian wave packet in a graphene sheet placed in a uniform and perpendicular magnetic field. We have observed that the temporal dynamics of the wave packet strongly depends on the initial pseudospin polarization. We have shown that direction of propagation of the wave packet can be controlled through pseudospin polarization.
\\
Further, we have studied the effect of pseudospin on trembling motion (ZB) of the Gaussian  wave packet in a graphene sheet placed in a quantizing magnetic field for non zero values of the initial momentum $k_{0x}$. Initially, these ZB oscillations  seemingly die out but reappear i.e. the amplitude of the ZB oscillations start to grow again without any fundamental frequency. So, for non-zero magnetic fields, ZB oscillations have a recurrent character and they do not die with time. The quantized (discrete) energy spectrum, which is a consequence of the magnetic field, is the main reason for the recurrent character of ZB  oscillations. This property is completely different from the zero magnetic field case, in which the energy spectrum is not quantized (discrete) and ZB of the wave packets  has a transient character.   \cite{transZB2007} Most importantly,  we have found that in the presence of a magnetic field, ZB  has strong dependence on initial pseudospin polarization, which can be regarded as pseudospinorial Zitterbewegung (PZB), a name given in a recent study. \cite{PZB}
\\
We have also studied wave packet dynamics in a one dimensional antidot lattice in graphene in the presence of the magnetic field. For this, we have  determined the Green's function matrices of the system both in frequency and time representations. From the poles of the Green's function, we have found that along the axis of antidot lattice, the energy spectrum is composed of Landau minibands with a unit shift at $K$ and $K^\prime $ points of graphene. In the time evolution of a wave packet, we find that wave packet dynamics are highly dependent on the initial pseudospin polarization; and the center of the Gaussian wave packet can be made to propagate along the antidot lattice by tuning the pseudospin parallel to the axis of antidot lattice. Also, when the strength of  the antidot potential is greater than the strength of the magnetic field, the wave packet becomes more confined in space  and the probability of finding it on the axis of antidot lattice significantly increases. Hence, we propose that quantum antidot channels can be made such that propagation of the wave packet can be controlled using pseudospin polarization and the strength of the antidot potential.
\\
Finally, we would like to point out that this work may lead to new insights for controlling currents  in both natural and artificial graphene systems by tuning initial pseudospin polarization, both in Landau quantized graphene and in a graphene antidot lattice in the presence of a magnetic field. This may lead to the preparation of graphene based nano gates  in which two different pseudospin polarizations perpendicular to each other can be used to get ``ON'' and ``OFF'' states.
\begin{appendices}
\appendix\section{Contour Integral}
To solve the second integral in  Eq. (\ref{eq:2.38}), let us represent the integral by $I$
\begin{multline} \label{C1}
I =  \frac{\al d \omega_g}{2\pi} \sum_{n^\prime  =-\infty}^\infty  \int_{-\infty}^\infty  d\eta \hspace{0.7mm}
 e^{-\textit i \omega_g\eta t}   \mathcal G^0(x_1;n^\prime  d;\eta)
\\
  \int_{-\frac{\pi}{d}}^{\frac{\pi}{d}} dp\hspace{2mm} e^{-\textit ipn^\prime  d}\left[ I_2-\al \hspace{1mm} \dot{\tilde{\mathcal G}}^0(p;0,0;\eta)\right]^{-1} \tilde{\mathcal G}^0(p;x_2;\eta).
\end{multline}
($I_2$ is unit matrix of order 2) where
\begin{equation*} \label{C2}
\eta = \frac{\omega}{\omega_g} \qquad  and \qquad d\omega = \omega_g d\eta
\end{equation*}
Note that in the above integral, each Green's function has a real pole at $\eta$=$\pm \sqrt{n}$. The poles of the term 
\begin{equation} \label{C3}
T_1 (\eta) = \left[ I_2 -\al \hspace{1mm} \dot{\tilde{\mathcal G}}^0(p;0,0;\eta)\right]^{-1} \tilde{\mathcal G}^0(p;x_2;\eta)
\end{equation}
can be seen to cancel  by reexpressing the Green's  functions  $\dot{\tilde{\mathcal G}}^0(p;0,0;\eta)$ and  $\tilde{\mathcal G}^0(p;x_2;\eta)$. For this purpose, by using Eqs. (\ref{eq:2.31a}) and (\ref{eq:2.32b}),  one can easily write $\al \dot{\tilde{\mathcal G}}^0(p;0,0;\eta)$ matrix as
\begin{equation} \label{C4}
\al \dot{\tilde{\mathcal G}}^0(p;0,0;\eta) = \frac{1}{\prod_{n=0}^\infty (\eta^2-n) }
\begin{pmatrix}
a_{11} &   a_{12}  \\
a_{12} &   a_{11}
\end{pmatrix}
\end{equation}
where we have defined 
\begin{equation*} \label{C10}
a_{11}(\eta)= \al_1 \eta \sum_{m=0}^\infty L_m \left[\frac{\omega^2_g l^2}{4\ga^2}\left(j\frac{d}{l} \right)^2\right] \prod_{n=0 \atop n\neq m}^\infty (\eta^2-n)
\end{equation*}
and
\begin{equation*} \label{C11}
a_{12}(\eta)=  \al_2\sum_{m=1}^\infty L_{m-1}^1 \left[\frac{\omega^2_g l^2}{4\ga^2}\left(j\frac{d}{l} \right)^2\right] \prod_{n=0 \atop n\neq m+1}^\infty (\eta^2-n)
\end{equation*}
with
\begin{equation*} \label{C5}
\al_1=  \frac{\al \omega_g}{4\pi\hbar\ga^2}\hspace{0.8mm} \hspace{0.8mm} \sum_{j=-\infty}^\infty e^{i pjd}  \hspace{0.8mm}   e^{-\frac{\omega^2_g l^2}{8\ga^2}\left(j\frac{d}{l} \right)^2} 		 	
\end{equation*}
and
\begin{equation*} \label{C12}	
\al_2= \frac{\textit i  \hspace{0.6mm} \al \omega_ g^2 l}{8\pi\hbar\ga^3} \hspace{0.8mm} \sum_{j=-\infty}^\infty  \left(j \frac{d}{l} \right) \hspace{0.8mm}  e^{i pjd}   \hspace{1mm}  e^{-\frac{\omega^2_g l^2}{8\ga^2}\left(j\frac{d}{l}\right)^2}.
\end{equation*}
Similarly, the $\tilde{\mathcal G}^0(p;x_2;\eta)$ matrix can be written as
\begin{equation} \label{C6}
\tilde{\mathcal G}^0(p;x_2;\eta)= \frac{1}{\prod_{n=0}^\infty (\eta^2-n) } 
\begin{pmatrix}
       b_{11}        & b_{12}  \\
    b_{12}        & b_{11}  

\end{pmatrix}
\end{equation}
where we have defined 
\begin{equation*} \label{C13}
b_{11}(\eta)=\beta_1\eta \sum_{m=0}^\infty L_m \left[\frac{\omega^2_g l^2}{4\ga^2}\left(j\frac{d}{l} \right)^2\right] \prod_{n=0\atop n\neq m}^\infty (\eta^2-n)    
\end{equation*}
and
\begin{equation*} \label{C14}
b_{12}(\eta)=  \beta_2 \sum_{m=1}^\infty L_{m-1}^1 \left[\frac{\omega^2_g}{4\ga^2 l^2}\left(j\frac{d}{l} \right)^2\right] \prod_{n=0\atop n\neq m+1}^\infty (\eta^2-n) ,
\end{equation*}
with
\begin{equation*} \label{C7}
\beta_1= \frac{ \omega_g}{4\pi\hbar\ga^2}\hspace{0.8mm} \hspace{0.8mm} \sum_{j=-\infty}^\infty e^{i pjd} \hspace{0.8mm}   e^{-\frac{\omega^2_g l^2}{8\ga^2}\left(\frac{jd-x_2}{l} \right)^2}
\end{equation*}
and 
\begin{equation*} \label{C15}		
\beta_2=\frac{\textit i  \hspace{0.6mm} \omega_g^2 l}{8\pi\hbar\ga^3} \hspace{0.8mm} \sum_{j=-\infty}^\infty   \left(\frac{jd-x_2}{l} \right) \hspace{0.8mm} e^{i pjd}  \hspace{1mm}  e^{-\frac{\omega^2_g l^2}{8\ga^2}\left(\frac{jd-x_2}{l}\right)^2}.
\end{equation*}
Note that in the above equations, $n$ is Landau index and $m$ is a dummy index for the Landau levels; the maximum value of $n$ and $m$ will be same.
\\
Finally substituting Eqs. (\ref{C4}) and (\ref{C6}) in  Eq. (\ref{C3}), the matrix $T_1$ will become
 \begin{multline*} 
T_1 (\eta) =  
 \Bigg[  
 \begin{pmatrix}
\prod_{n=0}^\infty (\eta^2-n)     	& 0 \\
0				& \prod_{n=0}^\infty (\eta^2-n)
\end{pmatrix} 
\\
- \al 
 \begin{pmatrix}
a_{11}(\eta)      	& a_{12}(\eta) \\
a_{12}(\eta) 	& a_{11}(\eta)
\end{pmatrix}
\Bigg]^{-1}  \hspace{1.4mm}
 \begin{pmatrix}
b_{11}(\eta)      	& b_{12}(\eta) \\
b_{12}(\eta) 	& b_{21}(\eta)  
\end{pmatrix}.
\end{multline*}  
In the above expression, the real poles ($\eta$=$\pm\sqrt{n}$)  cancel. The  above expression  can be solved  numerically. The   resultant $2 \times 2$ matrix with $c_{ij}(\eta)$ (where $i,j=1,2$) as matrix elements  can be written as (substituting $q=pd$)
\begin{equation} \label{C8}
Q (\eta) = \int_{-\pi}^{\pi} dq  e^{{-\textit iqn^\prime }}T_1 (\eta) = 
 \begin{pmatrix} 
c_{11}(\eta)      	& c_{12}(\eta)  \\
c_{21}(\eta) 	& c_{22}(\eta)  
\end{pmatrix}.
\end{equation}
Note that, the above integral can  be numerically calculated  by applying trapezoidal rule in the limits $-\pi$ to $\pi$ while keeping the trapezoidal step equal to $\frac{\pi}{10}$; this gives an accuracy up to five decimal points for each value of $n^\prime$.
\\
Putting  Eq. (\ref{C8}) in  Eq. (\ref{C1}), the integral $I$ becomes
\begin{equation} \label{C17}
I =  \frac{\al  \omega_g}{2\pi} \sum_{n^\prime  =-\infty}^\infty  \int_{-\infty}^\infty  d\eta \hspace{0.7mm}
 e^{-\textit i \omega_g\eta t}   \mathcal G^0(x_1,n^\prime  d;\eta) \hspace{0.5mm} . \hspace{0.5mm} Q(\eta).
\end{equation}
Using Eqs. (\ref{eq:2.29}) and (\ref{eq:2.30}), the matrix $ \al \mathcal G^0(x_1,n^\prime  d;\eta)$ can be written as
\begin{multline} \label{C18}
\al \mathcal G^0(x_1,n^\prime  d;\eta)= 
\begin{pmatrix}
       \ga_{1}\frac{1}{\eta}            & 0    \\
         0      &   \ga_{1} \frac{1}{\eta}
\end{pmatrix}
+  \sum_{n=1}^\infty
\\ \times
\begin{pmatrix}
       \ga_{1}     \eta
 \frac{ L_n \left[\Upsilon\right]}{\eta^2 -n}         & \ga_{2}  
 \frac{ L_{n-1}^1 \left[\Upsilon\right]}{\eta^2 -n}  \\
    \ga_{2}
 \frac{ L_{n-1}^1 \left[\Upsilon\right]}{\eta^2-n}         & \ga_{1}    \eta
 \frac{ L_n \left[\Upsilon\right]}{\eta^2 -n} 
\end{pmatrix}
\end{multline}
where $\Upsilon$, $\ga_1$ and $\ga_2$ are defined in Eqs. (\ref{eq:2.44}) and (\ref{eq:2.45}).
 Using the matrices $Q(\eta)$ and $ \al \mathcal G^0(x_1,n^\prime  d;\eta)$ in Eq. (\ref{C17}),  and breaking the matrix into two matrices, we get
\begin{multline} \label{C21}
I =  \frac{ \omega_g}{2\pi} \sum_{n^\prime  =-\infty}^\infty  \int_{-\infty}^\infty  d\eta \hspace{0.7mm}
 e^{-\textit i \omega_g\eta t} 
\begin{pmatrix}
 \ga_{1}\frac{1}{\eta} c_{11}(\eta)      	& \ga_{1}\frac{1}{\eta} c_{12}(\eta)  \\
 \ga_{1}\frac{1}{\eta} c_{21}(\eta) 	&  \ga_{1}\frac{1}{\eta} c_{22}(\eta)  
\end{pmatrix}
 \\ +
\frac{ \omega_g}{2\pi} \sum_{n^\prime  =-\infty}^\infty  \int_{-\infty}^\infty  d\eta \hspace{0.7mm}
 e^{-\textit i \omega_g\eta t} 
\begin{pmatrix}
   M_{11}(\eta) &    M_{12}(\eta) \\
    M_{21}(\eta) &    M_{22}(\eta)
\end{pmatrix} 
\end{multline}
(The first and second matrices correspond to $n$=$0$ and  $n>0$  Landau minibands, respectively) In above expression
\begin{multline*} 	
  M_{11 \atop 22}(\eta) =  \sum_{n=1}^\infty \Bigg( \ga_{1}     \eta
 \frac{ L_n \left[\Upsilon\right]}{\eta^2 -n} c_{11 \atop 22 }(\eta)
\\ 
+ \ga_{2}  
 \frac{ L_{n-1}^1 \left(\Upsilon\right)}{\eta^2 -n} c_{21 \atop 12}(\eta)\Bigg) ,
\end{multline*}
and
\begin{multline*}
  M_{12 \atop 21}(\eta) = \sum_{n=1}^\infty \Bigg( \ga_{1}      \eta
 \frac{ L_n \left[\Upsilon\right]}{\eta^2 -n} c_{12 \atop 21}(\eta)
\\ 
+ \ga_{2}  
 \frac{ L_{n-1}^1 \left[\Upsilon\right]}{\eta^2 -n} c_{22 \atop 11}(\eta) \Bigg).
\end{multline*}
In Eq. (\ref{C21}), the first matrix has a pole at $\eta$=$0$, while the second matrix has poles at $\eta$=$\pm\sqrt{n}$. We now use contour integration  with the Jordan lemma (closing the contour in the lower half plane for $t>0$) to evaluate the integrals. Results for  the two terms  in Eq. (\ref{C21}) are
\begin{equation}\label{C26}
\int_{-\infty}^\infty  d\eta \hspace{0.7mm} e^{-\textit i \omega_g\eta t}  \frac{c_{ij}(\eta)}{\eta}  =- i\pi\eta_+(t)  \left[c_{ij}(\eta)\right]_{\eta=0} ,
\end{equation}
\begin{multline} \label{C27}		
\int_{-\infty}^\infty  d\eta \hspace{0.2mm} e^{-\textit i \omega_g\eta t}   M_{11 \atop 22 }(\eta)  = -i \pi \eta_+(t) \sum_{n=1}^\infty    \cos \left(\frac{\omega_g \hspace{0.2mm} l}{\ga} \hspace{0.2mm} \frac{t}{\tau_o} \hspace{0.2mm} \sqrt n \right)
\\  \times
\left[ \ga_1 L_n \left[\Upsilon\right] c_{11 \atop 22}(\eta) +
\frac{\ga_2}{\sqrt{n}}          L_{n-1}^1 \left[\Upsilon\right] c_{21 \atop 12} (\eta) \right]_{\eta=\sqrt{n}}
\end{multline}
and 
\begin{multline} \label{C30}		
\int_{-\infty}^\infty  d\eta \hspace{0.2mm} e^{-\textit i \omega_g\eta t}   M_{12 \atop 21}(\eta)  =- \pi\eta_+(t)  \sum_{n=1}^\infty    \sin \left(\frac{\omega_g \hspace{0.2mm} l}{\ga} \hspace{0.2mm} \frac{t}{\tau_o} \hspace{0.2mm} \sqrt n \right) 
\\ \times
\left[ \ga_1 L_n \left[\Upsilon\right] c_{12 \atop 21}(\eta) +
  \frac{\ga_2}{\sqrt{n}}          L_{n-1}^1 \left[\Upsilon\right] c_{22 \atop 11} (\eta) \right]_{\eta=\sqrt{n}}.
\end{multline}
($\eta_+(t)$ is Heaviside unit step function and $i,j$=$1,2$). In the calculation of the expressions given by Eqs. (\ref{C26}),(\ref{C27})  and (\ref{C30}), we have  also used 
\begin{equation*}
\left[c_{11 \atop 22}(\eta)\right]_{\eta=-\sqrt{n}} = \left[c_{11 \atop 22}(\eta)\right]_{\eta=\sqrt{n}}
\end{equation*}
and
\begin{equation*}
\left[c_{12 \atop 21}(\eta)\right]_{\eta=-\sqrt{n}} = - \left[c_{12 \atop 21}(\eta)\right]_{\eta=\sqrt{n}}
\end{equation*}
which we found during the calculations.\\
Hence, Eq. (\ref{C26}) along with   Eqs. (\ref{C27})  and  (\ref{C30}) provide the complete solution for the integral $I$, which is the time representation of second term of the full Green's function $\mathcal G(x_1,x_2; t)_{K}$ given in Eq. (\ref{eq:2.38}).
\end{appendices}
\section*{Data Availability}
The data that support the findings of this study are available from the corresponding author upon reasonable request.
\vspace{2pt}
%
\end{document}